\documentclass[nofootinbib,prd,aps,twocolumn,preprintnumbers,amsmath,amssymb,superscriptaddress]{revtex4}
\usepackage{amsmath}
\usepackage{amssymb}
\usepackage{graphicx}
\usepackage{subfigure}
\usepackage{color}
\usepackage[colorlinks,linkcolor=magenta,anchorcolor=blue,citecolor=green]{hyperref}
\usepackage{ulem}
\usepackage{pifont}
\usepackage{makecell}
\begin{document}

\title{Mimetic Curvaton}

\author{Xin-zhe Zhang}
\email{zincz@hust.edu.cn}
\affiliation{School of Physics, Huazhong University of Science and Technology\\
Wuhan, 430074, China}

\author{Lei-Hua Liu}
\email{liuleihua8899@hotmail.com(co-corresponding author)}
\affiliation{Department of Physics, College of Physics,
	Mechanical and Electrical Engineering,
	Jishou University\\
	Jishou 416000, China}

\author{Taotao Qiu}
\email{qiutt@hust.edu.cn(corresponding author)}
\affiliation{School of Physics, Huazhong University of Science and Technology\\
Wuhan, 430074, China}

\begin{abstract}

    In this paper, we investigate the primordial perturbations of inflation model induced from the multi-field mimetic gravity, where there are two fields during inflation, and both adiabatic and isocurvature perturbation modes are generated. We show that although the original adiabatic perturbation mode indeed loses the kinetic term due to the constraint equation, by applying the curvaton mechanism where one of the fields is viewed as curvaton field, the adiabatic perturbation can be transferred from the isocurvature one at the end of inflation. Detailed calculations are performed for both inflationary and the consequent radiation-dominant and matter-dominant epochs. Therefore, the so-called ``non-propagating problem" of the adiabatic mode will not harm the multi-field mimetic inflation models. 
\end{abstract}

\maketitle

\section{Introduction}

The recently proposed mimetic gravity provides us with a novel mechanism of dark matter generation \cite{Chamseddine:2013kea, Golovnev:2013jxa, Sebastiani:2016ras}. Such kind of gravity introduces an auxiliary metric $\tilde{g}_{\mu\nu}$, which connects with our physical metric $g_{\mu\nu}$ via the conformal transformation
\begin{equation}
   g_{\mu\nu}=(\tilde{g}^{\alpha \beta}\partial_{\alpha}\phi \partial_{\beta}\phi)\tilde{g}_{\mu\nu}\ 
    \label{conformal transformation}
\end{equation}
where $\phi$ is the auxiliary field. By varying the Hilbert-Einstein action with respect to the auxiliary metric $\tilde{ g}^{ \mu \nu}$, one can have an additional term in $T_{\mu\nu}$ that is proportional to $(G-T)$ (the trace difference between $G_{\mu\nu}$ and $T_{\mu\nu}$), mimicking the cold dark matter according to the fact that the effective pressure caused by this term is zero. This means that due to such a transformation, the longitudinal mode of gravity now become dynamical. Moreover, from \eqref{conformal transformation} one can find that the scalar field $\phi$ obeys the constraint of
\begin{equation}
    g^{\mu\nu}\partial_\mu\phi\partial_\nu\phi=-1~.
    \label{constraint}
\end{equation}
thus a Lagrangian multiplier term can be added into Hilbert-Einstein action so that the physical metric $g^{ \mu \nu}$ is still considered as variable instead of auxiliary metric $\tilde{ g}^{ \mu \nu}$ with no change on equations of motion \cite{Chamseddine:2014vna}.

The mimetic gravity can also be extended to describe inflation where a potential $V(\phi)$ of the scalar is introduced as the usual single field inflation model. However, as in \cite{Chamseddine:2014vna} the authors have pointed out, considering the constraint equation \eqref{constraint}, one actually cannot get a wavelength-dependent solution. In other words, we cannot define the quantum state of the perturbations in usual way. Therefore we need to modify the theory by adding new degree of freedoms. One way of doing this is to add the higher derivative terms such as $\Box\phi$ to the action, but pure $\Box\phi$ will generally lead to pathology such as gradient instabilities \cite{Chamseddine:2014vna, Ramazanov:2016xhp, Ijjas:2016pad, Firouzjahi:2017txv} (see also \cite{Takahashi:2017pje} for more complicated case). On the other hand, in order to get a stable mimetic gravity theory, various ways of modifying the higher derivative terms have been discussed, see e.g., \cite{Chaichian:2014qba, Nojiri:2014zqa, Cognola:2016gjy, Hirano:2017zox, Zheng:2017qfs, Gorji:2017cai, Casalino:2018tcd, Casalino:2018wnc, HosseiniMansoori:2020mxj}.  Another way is to add extra fields, by which one then gets a multi-scalar mimetic gravity \cite{Firouzjahi:2018xob, Mansoori:2021fjd}. While Ref. \cite{Firouzjahi:2018xob} considered two fields with a flat metric $\delta_{ab}$ in field space, in \cite{Mansoori:2021fjd} the authors  extended it into an arbitrary metric $G_{ab}$, violating the shift symmetry of the fields. By decomposing perturbations into adiabatic and entropy modes as the usual approach performed on the multi-field inflation models, one can calculate their evolution behaviors respectively. However, Ref. \cite{Mansoori:2021fjd} claimed that, due to the constraint equation, the velocity of adiabatic mode can be described as a function of both two modes, eliminating the kinetic term of adiabatic mode in the total action, which means that the adiabatic mode does not propagate. Moreover, this conclusion holds independence on gauge choice. In this case, it will bring problems because the adiabatic perturbation of each space-time point will evolve independently, and the global homogeneity of our Universe required by the Cosmic Microwave Background (CMB) observations cannot be guaranteed. Therefore there must be a mechanism to avoid the case. 

One feasible mechanism is the so-called curvaton mechanism \cite{Lyth:2001nq,Lyth:2002my}, in which it states that there is an extra light scalar field (comparing with inflaton) responding to the curvature perturbation. Since the curvaton field is much lighter than the inflaton field, its effects on background trajectory can be neglected, meanwhile, the curvaton will generate the most of the perturbations which are isocurvature perturbations. At the end of inflation, the curvaton field will decay into radiation or matter, by which it will transfer the isocurvature perturbation into the curvature perturbation. Moreover, we can realize the curvaton mechanism from the inflaton decay that can be dubbed as one field inflationary theory \cite{Liu:2019xhn}. 
In this paper, we consider the curvaton mechanism under the framework of multi-field mimetic inflation, where one of the fields acts as a curvaton, thus one can still get the adiabatic perturbations after inflation \cite{Liu:2020zzv}. Inspired by this framework, the adiabatic perturbation is sourced by the isocurvature one which is propagating, therefore the ``non-propagating" problem demonstrated above can be avoided.

This paper is organized as follows: in Sec. II we briefly review the mimetic inflation model (both single- and multi-field), and show the problems for both cases; in Sec. III we study the background evolution for our mimetic curvaton model; in Sec. IV we investigate the perturbations in our model, by using the curvaton mechanism, for both inflationary and matter-dominant epochs. The final curvature perturbation in matter-dominant epoch is also obtained. Sec. V gives the main conclusion and final discussion.

\section{Mimetic gravity for inflation}
\label{mimetic gravity}
The original single-field inflation model that arises from mimetic gravity is described by the action \cite{Chamseddine:2014vna}: 
\begin{align}
    S&= \int d^ 4 x \sqrt{ -g}\bigg[ \frac{ M_{ P}^{ 2}}{ 2} R + \lambda \left( g^{ \mu \nu} \partial_ \mu \phi \partial_ \nu \phi + M^4 \right) \nonumber\\
    &- V( \phi) \bigg]\ ,
    \label{single action}
\end{align}
where the constraint is put inside the action with a multiplier $\lambda$ and the potential $V(\phi)$ is also added to the action. Considering the dimension, the constraint term is written as $g^{\mu\nu}\partial_\mu\phi\partial_\nu\phi+M^4$ where $M$ corresponds to the typical energy scale of $\phi$. One can always do the rescaling of $\phi$ as $\phi\rightarrow\phi/M$ to reduce the constraint to the form of \eqref{constraint}. 

From action \eqref{single action}, it is straightforward to obtain the Friedmann equations in flat Friedmann-Lema\^{i}tre-Robertson-Walker (FLRW) metric:
\begin{align}
    \begin{cases}
    3 M_{ P}^{ 2} H^{ 2}&= V - 2 \lambda M^4 \ ,\\
    M_{ P}^{ 2} \Dot{ H}&= \lambda M^4 \ ,
    \end{cases}
    \label{Friedmann}
\end{align}
where $\partial^\mu\phi\partial_\mu\phi$ is eliminated by the constraint thus does not appear as in the usual case. As can be seen, $\lambda$ appears in Friedmann equations as a part of energy density and totally determines the time evolution of Hubble parameter. From the Friedmann equations, one gets the equation of motion of $\lambda$ as
\begin{align}
    \Dot{ \lambda} + 3 H \lambda - \frac{ \Dot{ V}}{ 2M^4} = 0\ ,
    \label{equation lambda}
\end{align}
and its solution is
\begin{align}
    \lambda= \frac{ 1}{ 2M^4} a^{ -3} \int_{t_i}^t dt^ \prime \Dot{ V} a^{ 3} + C_{ 1} a^{ -3}\ ,
    \label{solution for lambda}
\end{align}
where $C_ 1$ is an integral constant. Thus the parameter $\lambda$ is determined by potential $V$ and boundary condition. From the above solution one can obtain remarks as follows:\\
1) during slow-roll inflation ($t_i<t<t_e$) where slow-roll parameter $\varepsilon \equiv - \Dot{ H} / H^ 2 \simeq 0$, one has $\lambda = M_ P^ 2 \Dot{ H}/M^4 \simeq 0$, therefore $V = 3 M_ P^ 2 H^ 2$ which is almost independent on time, namely $\int_{t_i}^t dt \Dot{ V} a^ 3 \simeq 0$ till $t=t_e$. These in turn implies that $C_ 1$ vanishes;\\
2) from the end of inflation to the beginning of matter-dominance ($t_e<t<t_{mi}$) where needs $\varepsilon \simeq 1$ to terminate the inflation, from \eqref{Friedmann} we have $V=-\lambda M^4$. Taking $C_1=0$, this leads to $V \propto a^{ -2}$;\\
3) from matter-dominance to now ($t_{mi}<t<t_0$) where mimetic field $\phi$ has almost totally decayed, yielding $V \simeq 0$ and $\dot V\simeq 0$, we have $\int_{t_{mi}}^t dt \Dot{ V} a^ 3 =0$ till $t=t_0$, so $\int_{t_i}^t dt \Dot{ V} a^ 3 =\int_{t_e}^{t_{mi}} dt \Dot{ V} a^ 3$ is a constant (definite integration), making $\rho = V - 2 \lambda M^4 \propto a^{ -3}$ as unrelativistic matter. This is why we can consider it as a candidate of dark matter. Setting $C_2\equiv\int_{t_e}^{t_{mi}} dt \Dot{ V} a^ 3$, we will find that $C_2$ is nothing but the dark matter energy density today,
\begin{align}
    \rho_{ m 0} = \rho_ m a^ 3 = - \int_{ t_ i}^{t_0} \Dot{ V} a^ 3 dt = \int_{t_e}^{t_{mi}} dt \Dot{ V} a^ 3=- C_ 2\ ,
\end{align}
therefore the constant can be treated as a connection between the early epoch (right after inflation) and the late epoch of our Universe. It indicates that the cold dark matter fraction in mimetic gravity is produced in preheating epoch.


However, the single-field mimetic gravity cannot describe inflation properly. Due to the constraint equation, the universal equation of motion for the scalar perturbation $\delta\phi$ is obtained in Ref. \cite{Chamseddine:2014vna}:
\begin{align}
    \delta \Ddot \phi + H \delta \Dot \phi + ... = 0 \ ,
    \label{problem of single mimetic}
\end{align}
where there is no spatial derivative, thus $\delta \varphi$ is valid irrelevant of the wavelength. Note that we have omitted parts of Eq. \eqref{problem of single mimetic} which is not important for our discussion. Therefore we have problem of defining the quantum origin of the perturbations. To be precise, from Eq. \eqref{problem of single mimetic} we cannot have a plane-wave like solution as usual, and the momentum of the quantum fluctuations is not well-defined, either.
In order to avoid this problem, in Ref. \cite{Mansoori:2021fjd} the model is extended by adding another mimetic field to the system. If the mimetic gravity is described by more than one field with transformation $g_{ \mu \nu}=( G_{ a b} \Tilde{ g}^{ \alpha \beta} \partial_ \alpha \phi^ a \partial_ \beta \phi^ b) \Tilde{ g}_{ \mu \nu}$, the action becomes
\begin{align}
    S &= \int d^ 4 x \sqrt{ -g}\bigg[ \frac{ M_{ P}^{ 2}}{ 2} R + \lambda \left( G_{ a b} g^{ \mu \nu} \partial_ \mu \phi^ a \partial_ \nu \phi^ b + M^4 \right) \nonumber\\
    &- V( \phi^ a) \bigg]\ ,
    \label{multiple action}
\end{align}
where $G_{ a b}$ is metric of field space. As discussed in \cite{Mansoori:2021fjd} as well as in \cite{Shen:2019nyp}, the model indeed has two scalar degrees of freedom as the same as in normal double-field inflation models. For background evolution, Eqs. \eqref{Friedmann}-\eqref{solution for lambda} are still applicable. The authors have calculated the perturbations generated by this model using the well-known adiabatic-entropy decomposition, finding that different from normal double-field inflation models, the adiabatic mode perturbation does not propagate in multi-field mimetic gravity! This problem occurs due to the fact that the perturbed mimetic constraint combines the adiabatic mode perturbation and the entropy mode perturbation, which gives the relation
\begin{align}
    \Dot{ u}_ T = \varepsilon H u_ T + \Dot{ \theta} u_ N\ ,
    \label{relationship adiabatic entropy}
\end{align}
where $u_ T$ and $u_ N$ are adiabatic mode and entropy mode perturbations respectively, $\varepsilon$ is slow-roll parameter, $\theta$ is the angle of the tangent to the background trajectory with respect to one of the axis in the field space (see \cite{Gordon:2000hv, Mansoori:2021fjd}). As a result, the perturbed action becomes \cite{Mansoori:2021fjd}
\begin{align}
    &\delta^{ (2)} S = \int d^ 4 x a^ 3 M_ P^ 2 \varepsilon H^ 2 \Bigg[ \mathcal{ L}_{ u_ N} + \mathcal{ L}_{ u_ T} + 2 \text{sgn}( \pm 1) \dot{ \theta} u_ T \nonumber\\
    &\times \dot{ u}_ N - 2 \dot{ \theta} u_ N \dot{ u}_ T - 2 \Bigg( \frac{ M_{ NT}^ 2}{ M_ P^ 2 \varepsilon H^ 2} - \varepsilon H \dot{ \theta} \Bigg) u_ T u_ N + \frac{ 2}{ M_ P^ 2 H} \nonumber\\
    &\times \delta \lambda \Bigg( u_ T - \frac{ \dot{ u}_ T}{ \varepsilon H} + \frac{ 1}{ \varepsilon H} \dot{ \theta} u_ N \Bigg) \Bigg] \ ,
    \label{seconde order perturbed action}
\end{align}
with
\begin{align}
    \mathcal{L}_{ u_ N} = \text{sgn}( \pm 1) \left( \dot{ u}_ N^ 2 - \frac{ 1}{ a^ 2} \dot{ u}_ N^ 2 \right) + \left( \dot{ \theta}^ 2 - \frac{ M_{ NN}^ 2}{M_ P^ 2 \varepsilon H^ 2} \right) u_ N^ 2
    \label{lagrangian entropy mode}
\end{align}
and
\begin{align}
    \mathcal{L}_{ u_ T} = \dot{ u}_ T^ 2 - \frac{ 1}{ a^ 2} \left( \partial u_ T \right)^ 2 - 2 \varepsilon H u_ T \dot{ u}_ T \nonumber\\
    + \left( \text{sgn}( \pm 1) \dot{ \theta}^ 2 - \frac{ M_{ TT}^ 2}{ M_ P^ 2 \varepsilon H^ 2} \right) u_ T^ 2
    \label{lagrangian adiabatic mode}
\end{align}
where 

\begin{align}
    M_{ NN}^ 2 &\equiv \frac{ 1}{ 2}( N^ a N^ b V_{ ;ab} - \text{sgn}( \pm 1) \varepsilon H^ 2 \mathbb{ R},\nonumber\\
    M_{ TT}^ 2 &\equiv \frac{ 1}{ 2} T^ a T^ b V_{ ;ab} + \varepsilon H( T^ a V_{ ;a} + M_ P^ 2 \varepsilon H^ 3( 3 - \varepsilon)),\nonumber\\
    M_{ TN}^ 2 &\equiv \frac{ 1}{ 2}( T^ a N^ b V_{ :ab} + \varepsilon H N^ a V_{ ;a}),
\end{align}
where $T^a$ and $N^a$ are the tengent and normal unit vectors with respect to the background trajectory, while $\mathbb{R}$ is the Ricci scalar of field space. Substituting Eq. \eqref{relationship adiabatic entropy} into $\mathcal{L}_{ u_ T}$ in \eqref{lagrangian adiabatic mode}, it appears that $\dot{ u}_T^2$ can be re-expressed using $u_T$ and $u_N$, thus the kinetic term of $u_T$ get lost. Moreover, this problem cannot be avoided by gauge transformations. In the next section, we will involve curvaton mechanism to solve this problem.

\section{Curvaton Mechanism in Multi-field Mimetic gravity}
\label{curvaton mechanism}
In this section, we will investigate the curvaton mechanism in multi-field mimetic gravity. In order to show the realization of curvaton mechanism, with the freedom of choosing the metric of field space $G_{ab}$, we choose the following metric:
\begin{align}
    G_{ab}=diag\left\{1,6\sinh{}^2\left( \frac{ \varphi}{ \sqrt{ 6}M} \right)\right\}~.
\end{align}
This metric is also applied to the so-called $\alpha$-attractor inflation models \cite{Kallosh:2013pby, Kallosh:2013lkr, Kallosh:2013hoa, Kallosh:2013maa, Kallosh:2013daa}, which can preserve the local conformal symmetry with the transformation $\Tilde{ g}_{ \mu \nu} = \exp[ -2 \sigma( x)] g_{ \mu \nu}$, $\Tilde{ \chi} = \exp[ \sigma( x)] \chi$ and $\Tilde{ \phi}^ a= \exp[ \sigma( x)] \phi^ a$ with the original inflaton fields $\chi$ and $\phi^ a$. Moreover, it can avoid the intitial condition problem for inflationary potential as well.
Applying this metric, action (\ref{multiple action}) becomes
\begin{align}
    S &= \int d^ 4 x \sqrt{ -g}\Bigg \{ \frac{ M_ P^ 2}{ 2} R + \lambda\Bigg[ - \frac{ 1}{ 2} \Dot{ \varphi} + \frac{ 1}{ 2 a^ 2} \partial_ i \varphi \partial^ i \varphi \nonumber\\
    &+ 3 \sinh^ 2 \left( \frac{ \varphi}{ \sqrt{ 6}M} \right) \left( - \Dot{ \theta}^ 2 + \frac{ 1}{ a^ 2} \partial_ i \theta \partial^ i \theta \right) + M^4 \Bigg] \nonumber\\
    &- V( \varphi, \theta)\Bigg\}\ .
    \label{action multiple field mimetic inflation}
\end{align}
The equations of motion of this two mimetic fields are
\begin{align}
    \begin{cases}
        \Ddot{ \varphi} + 3 H \Dot{ \varphi} + \frac{ \dot{ \lambda}}{ \lambda} \dot{ \varphi} - \frac{ 1}{ a^ 2} \partial_ i^ 2 \varphi + \frac{\sqrt{ 6}}{M} \sinh \left( \frac{ \varphi}{ \sqrt{ 6}M} \right)\\
        \times \cosh \left( \frac{ \varphi}{ \sqrt{ 6}M} \right) \left( - \dot{ \theta}^ 2 +
        \frac{ 1}{ a^ 2} \partial_ i \theta \partial^ i \theta \right) + \frac{ 1}{ \lambda} \frac{ \partial V \left( \varphi, \theta \right)}{ \partial \varphi} = 0 \ ,\\
        \\
        \Ddot{ \theta} + 3 H \dot{ \theta} + \frac{ \dot{ \lambda}}{ \lambda} \dot{ \theta} + \sqrt{ \frac{ 2}{ 3M^2}} \coth \left( \frac{ \varphi}{ \sqrt{ 6}M} \right) \dot{ \varphi} \dot{ \theta} - \partial_ i^ 2 \theta\\
        - \sqrt{ \frac{ 2}{ 3M^2}} \coth \left( \frac{ \varphi}{ \sqrt{ 6}M} \right) \partial_ i \varphi \partial^ i \theta + \frac{ 1}{ \lambda} \frac{ \partial V \left( \varphi, \theta \right)}{ \partial \theta} = 0 \ ,
    \end{cases}
    \label{explicit eom}
\end{align}
while the mimetic constraint is
\begin{align}
    \dot{\varphi}^2-\frac{(\partial\varphi)^2}{a^2}+6\sinh^2\left(\frac{\varphi}{\sqrt{6}M}\right)[\dot{\theta}^2-\frac{1}{a^2}(\partial\theta)^2]=2M^4\ .
    \label{mimetic constraint multiple field}
\end{align}
In the following, we will analyze the background evolution at two epoches separately, namely the inflation epoch and the following matter-dominant epoch. For the sake of simplicity, we will ignore the reheating and radiation-dominant epoch, assuming that these epoches took place instantaneously. 
\subsection{the inflation epoch}
\label{inflation background}
\begin{figure}[t]
    \centering
    \subfigure{\includegraphics[width=3.in]{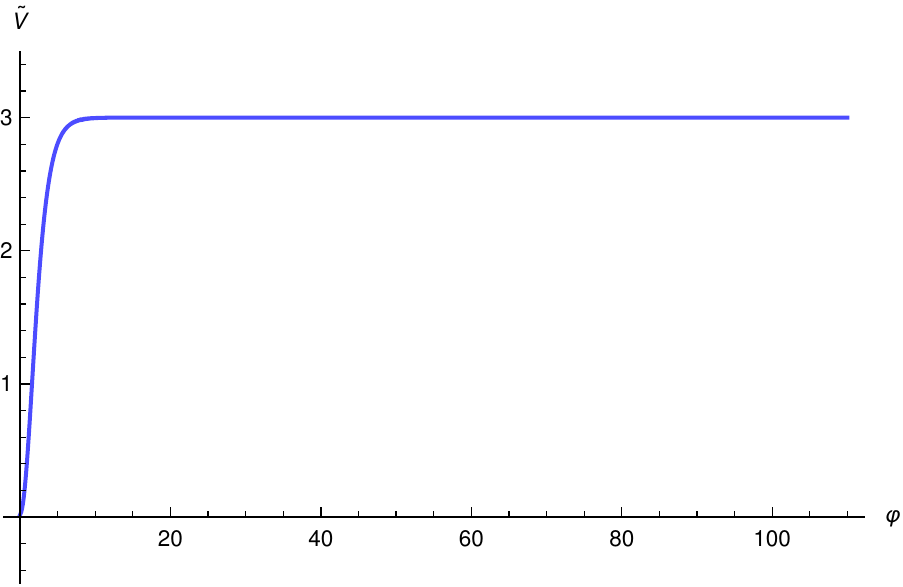}}
    \caption{The re-scaled potential $\tilde{ V}( \varphi) \equiv V( \varphi) / M_ P^ 2 H_ i^ 2$, which is flat at large $\varphi$ region and stops inflation along with the decreasing of $\varphi$.}
    \label{potential plot}
\end{figure}

In this subsection, we would like to show the realization of a inflation era in our model. We regard $\varphi$ as inflaton which drives inflation, while $\theta$ as curvaton, who plays little role in the background level, but plays main role in creating the curvature perturbation. As demonstrated in the original curvaton paper \cite{Lyth:2001nq}, we require that the potential along with the trajectory of $\theta$ should be sufficiently flat, namely $| \partial^ 2 V/ \partial \theta^ 2| \ll H^ 2$. One of the convenient choice is to consider a potential that is nearly independent of $\theta$, namely $V( \tanh( \varphi/ \sqrt{ 6}M), \theta) \simeq V( \varphi)$. Since the background behavior at this epoch is mainly determined by the inflaton $\varphi$, the equations of $\varphi_ 0$ and $\theta_0$ in Eq. \eqref{explicit eom} become:
\begin{align}
    \Ddot{ \varphi}_0 + 3 H \Dot{ \varphi}_0 + \frac{ \dot{ \lambda}}{ \lambda} \dot{ \varphi}_ 0 - \sqrt{ \frac{ 1}{ 6M^2}} \left( 2M^4 - \Dot{ \varphi}_0^ 2 \right) + \frac{ 1}{ \lambda} \frac{ d V}{ d \varphi_ 0} \simeq 0\ ,
    \label{equation of motion of varphi}
\end{align}
\begin{align}
    \Ddot{\theta}_0 + 3 H \Dot{ \theta}_ 0 + \frac{ \dot{ \lambda}}{ \lambda} \dot{ \theta}_ 0 + \sqrt{ \frac{ 2}{ 3M^2}} \coth \left( \frac{ \varphi_0}{ \sqrt{ 6}M} \right) \Dot{ \varphi}_ 0 \Dot{ \theta}_ 0 \simeq 0\ .
    \label{equation of motion of theta}
\end{align}
As the Refs. \cite{Kallosh:2013pby, Kallosh:2013lkr, Kallosh:2013hoa, Kallosh:2013maa, Kallosh:2013daa} has suggested, the specific form of $V(\varphi)$ can be chosen as $V( \varphi) \propto \text{ tanh}^{ 2 n}( \varphi/ \sqrt{ 6}M)$ at large $\varphi$ plotted in Fig. \ref{potential plot}. In Fig. \ref{potential plot} one can see that, the potential is rather flat in the large field stage, which is reasonable to cause a period of slow-roll inflation era. 

From the Friedmann equation \eqref{Friedmann} as well as the equation for $\lambda$ \eqref{equation lambda}, the Hubble parameter can be directly expressed as the potential functional $V(\varphi)$. However, to know its behavior with respect to time variable such as $t$ or $a$, we have to know the form of $\varphi(t)$ as well. 

\begin{figure}[t]
    \centering
    \subfigure{\includegraphics[width=3.in]{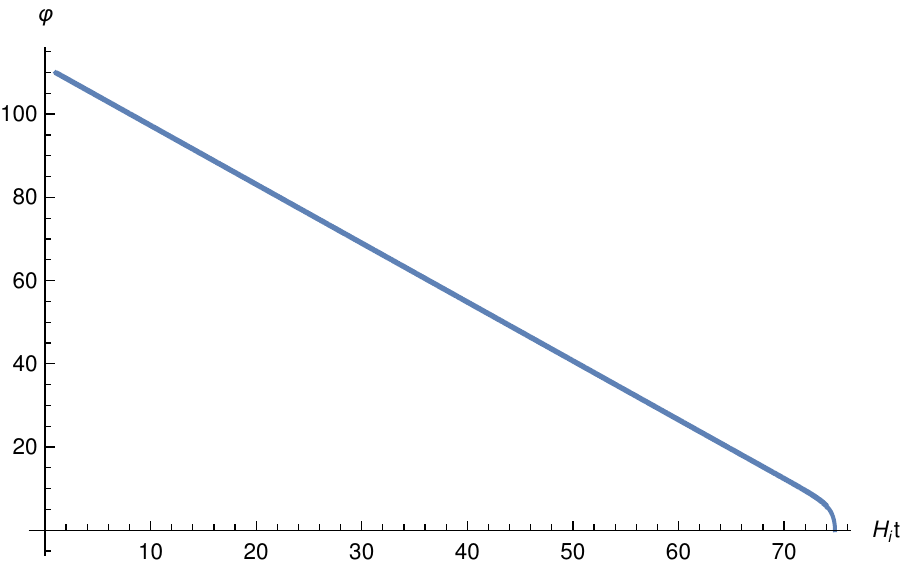}}
    \caption{The numerical plot of $\varphi$ according to Eq. \eqref{equation of motion of varphi}. The value of $\varphi$ is (nearly) monotonically decreasing, with the slope being its time derivative: $\dot\varphi\simeq-\sqrt{2}M^2$.}
    \label{varphi plot}
\end{figure}

It is not difficult to get one solution as $\theta_ 0 \simeq 0$ with the sufficiently flatness of potential along with the trajectory of $\theta$. Moreover, in the current case we can also utilize the constraint equation \eqref{mimetic constraint multiple field} without solving the equation of motion of $\varphi$ as usual inflation model. After performing slow-roll conditions and neglecting spatial derivative term as well as the solution of $\theta_0$ field, Eq. \eqref{mimetic constraint multiple field} gives
\begin{align}
    \dot\varphi\simeq \pm\sqrt{2}M^2~.
    \label{dotvarphi}
\end{align}
As a consistency check, we also numerically solve the equation of motion \eqref{equation of motion of varphi}, and plot the behavior of $\varphi(t)$ in Fig. \ref{varphi plot}. while the slope gives the value of $\dot\varphi$. One can see from the plot that Eq. \eqref{dotvarphi} is indeed the solution of the equation of motion \eqref{equation of motion of varphi} at large field value region. 

\begin{figure}[t]
    \centering
    \subfigure{\includegraphics[width=3.in]{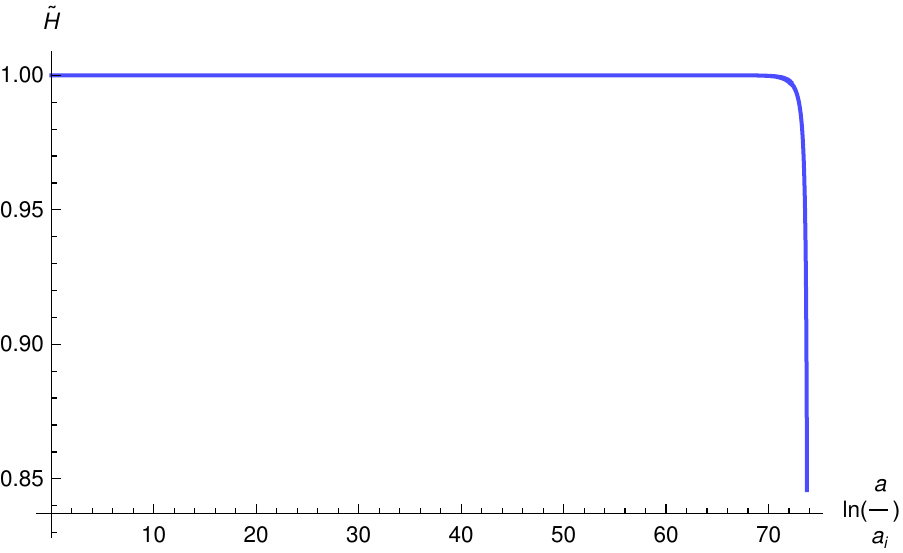}}
    \subfigure{\includegraphics[width=3.in]{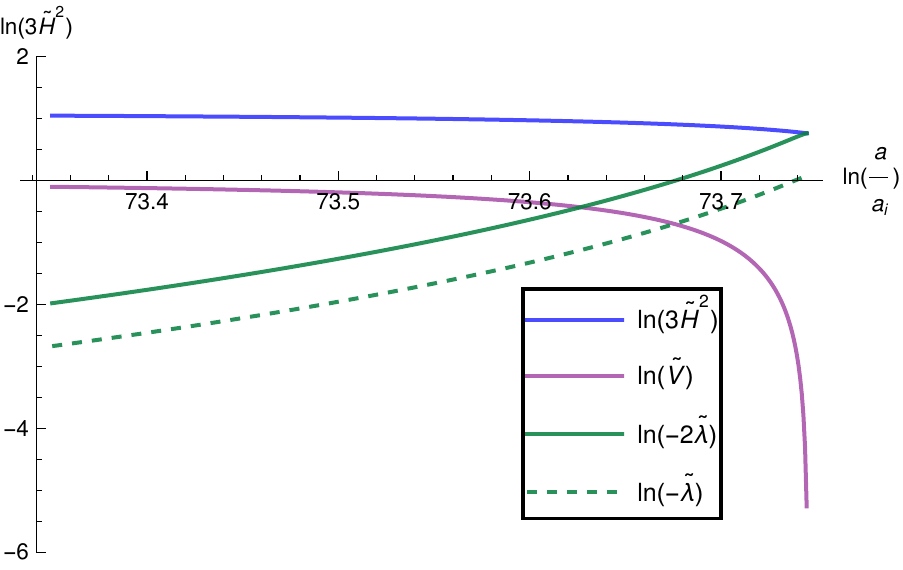}}
    \caption{The top figure shows the re-scaled Hubble parameter $\tilde{ H} \equiv H / H_ i$ in inflation with $H_ i$ as the initial value of Hubble parameter in inflationary epoch. To satisfy the requirement of slow-roll assumption, $H$ should be flat until $a / a_ i \geq \exp( 67)$. The bottom figure shows the evolution of re-scaled energy density $\tilde{ \rho} \equiv 3 \tilde{ H}^ 2 = \tilde{ V} - 2 \tilde{ \lambda}$ in blue with $\tilde{ V} \equiv V / M_ P^ 2 H_ i^ 2$ in purple and $- 2 \tilde{ \lambda} \equiv - 2 \lambda M^4 / M_ P^ 2 H_ i^ 2$ in green solid line. In order to terminate inflation, there should be $V = - \lambda M^4$ which is showed in green dashed line.}
    \label{background plot}
\end{figure}

Making use of the above result, one can get the numerical plot of the square of Hubble parameter $H^ 2$ as in Fig. \ref{background plot}. We show that the Hubble parameter is flat at the beginning and then decreases, which are correspondingly the inflation era and its end. We also plot the evolution behavior of each component (say, $V$ and $\lambda M^4$) of $H^ 2$. Following the analysis in the previous section, the inflation ends when $V=-\lambda M^4$, namely $\ln(-2\lambda M^4)=\ln(V)+\ln 2$. As showed in the Fig. \ref{background plot}, it actually happens at the time when the e-folding number $N\equiv \ln(a/a_i)\simeq 73.68$, which is consistent with the current constraints on inflation.

\subsection{matter-dominated epoch}
\label{MD background}

In the matter-dominated epoch, the inflaton field $\varphi$ has rolled along the potential from the large field region to the small field region. Therefore we have
\begin{align}
    \frac{\varphi_ 0}{M} \ll 1\ ,\ 
    \sinh \left( \frac{ \varphi_0}{ \sqrt{ 6}M} \right) \simeq \frac{ \varphi_ 0}{ \sqrt{ 6}M}\ ,\
    \cosh \left( \frac{ \varphi_0}{ \sqrt{ 6}M} \right) \simeq 1 \ .
    \label{assumptions md}
\end{align}
Moreover, according to Eq. \eqref{solution for lambda} and the analysis in Sec. \ref{mimetic gravity}, one has
\begin{align}
    \lambda \propto a^{ -3}\ .
    \label{lambda md}
\end{align}
The equation of $\varphi_ 0$ in Eq. \eqref{explicit eom} becomes
\begin{align}
    \Ddot{ \varphi}_ 0 - \frac{ 1}{ \varphi_ 0}( 2 M^4- \dot{ \varphi}_ 0^ 2) + \frac{ 1}{ \lambda} V^{ \prime} = 0\ .
    \label{eom varphi md}
\end{align}
Note that in this case, the $3H$ friction term that is to appear in the equation of motion has been cancelled by $\Dot{ \lambda} / \lambda$ term, so the equation looks like that in Minkowski space time. That means, the expansion effect is somehow ``screened" by the multiplier $\lambda$ with the behavior \eqref{lambda md}.

In order to maintain the matter-dominated epoch, the potential needs to quit from its flatness, and the most natural choice is that it behaves like a mass squared potential, at the bottom of which the field is oscillating. In usual case, the field will behave like ordinary matter as well \cite{Turner:1983he}. However, the $-2 M ^ 4 / \varphi_ 0$ term in Eq. \eqref{eom varphi md} will break such a behavior. To avoid this, we need to construct the potential to be $V( \varphi) \simeq \lambda [ m_{ eff}^ 2 \varphi_ 0^ 2 + \alpha \ln( \varphi_ 0 / \varphi_ 1)]$ at $\varphi_ 0 / M \ll 1$, with $m_{eff}$ as the effective mass of the inflaton field in matter domination and $\varphi_ 1$ as a constant comparable with $\varphi_ 0$. The last term with $\alpha = 2 M^ 4$ is delicately chosen to compensate $-2 / \varphi_ 0$ in the second term of Eq. \eqref{eom varphi md} without considering any physical origin since we purpose on the realization of mimetic curvaton. The solution of $\varphi_ 0$ can be obtained as
\begin{align}
    \varphi_ 0^{ 2} = A \sin \left( m_{ eff} t \right)\ .
    \label{solution varphi md}
\end{align}
From the Friedmann equation \eqref{Friedmann} we see that, in order to avoid the potential energy dominating the universe, we require $V \ll -2 \lambda M^4$, which will give rise to $m_{eff}^2A\ll 1$ constraining on the effective mass of $\varphi$ (modulo the amplitude $A$, which can be obtained from previous process of the inflaton's evolution.) Moreover, if we furtherly constrain the $m_{eff}$ to be $m_{ eff} t \simeq m_{ eff} / H \ll 1$, we can then pick up the leading order of the Taylor expansion of \eqref{solution varphi md}, which gives
\begin{align}
   \varphi_ 0^{ 2} \simeq A m_{ eff} t \ .
   \label{solution varphi md 1}
\end{align}


Moreover, under the assumptions \eqref{assumptions md} and solution \eqref{solution varphi md 1}, the \eqref{explicit eom} becomes
\begin{align}
    \Ddot{ \theta}_ 0 + 2 \frac{ \dot{ \varphi}_ 0}{ \varphi_ 0} \dot{ \theta}_ 0 = 0\ ,
    \label{eom theta md}
\end{align}
and it is easy to get the solution
\begin{align}
    \dot{ \theta}_ 0 = \frac{ C_ \theta}{ \varphi_ 0^ 2} \simeq \frac{ C_ \theta}{ A m_{ eff} t}\ .
    \label{solution theta md}
\end{align}
where $C_ \theta$ is a integral constant. On the other hand, the mimetic constraint \eqref{mimetic constraint multiple field} becomes
\begin{align}
    \frac{ A m_{ eff}}{ 4 t} + \frac{ C_ \theta^ 2}{ M^ 2 A m_{ eff} t} = 2 M^ 4\ .
\end{align}
The constraint seems to be inconsistent with the solution \eqref{solution varphi md} and \eqref{solution theta md}. The inconsistency arises because as the inflation has ended, the inflaton field (maybe as well as the curvaton field) will decay into other products. This will make the constraint equation no longer the form of \eqref{mimetic constraint multiple field}, but should also include the decay products. We denote the decaying products as another field $\psi$, and the constraint equation should be $\dot{ \varphi}^ 2 + 6 \sinh^ 2( \varphi / \sqrt{ 6}) \dot{ \theta}^ 2 + \dot{ \psi}^ 2 = 2M^4$. The detailed calculation will be modified accordingly as can be seen in the Appendix \ref{app}, and our main result will not get changed.

\section{Generating Curvature perturbations}
\label{curvature perturbation}

In this section, we will investigate the perturbations generated from the curvaton model, especially how adiabatic perturbations can be transferred from the isocurvature ones. 
We perturb the two scalar fields as $\varphi = \varphi_ 0 + \delta \varphi$ and $\theta = \theta_ 0 + \delta \theta$. Therefore according to the total equations of motion \eqref{explicit eom}, the explicit perturbed equations of motion are
\begin{align}
    \begin{cases}
    \delta \Ddot{ \varphi} + 3 H \delta \Dot{ \varphi} + \frac{ \Dot{ \lambda}}{ \lambda} \delta \Dot{ \varphi} - \frac{ 1}{ a^ 2} \partial_ i^ 2 \delta \varphi -\frac{1}{M^2} \Big[ \sinh^ 2 \left( \frac{ \varphi_ 0}{ \sqrt{ 6}M} \right)\\
    - \cosh^ 2 \left( \frac{ \varphi_ 0}{ \sqrt{ 6}M} \right)\Big]  \left( \Dot{ \theta}_ 0^ 2 - \frac{ 1}{ a^ 2} \partial_ i \theta_ 0 \partial^ i \theta_ 0 \right)\delta \varphi \\
    - \sqrt{ \frac{ 2}{ 3M^2}} \sinh \left( \frac{ \varphi_ 0}{ \sqrt{ 6}M} \right)     \cosh \left( \frac{ \varphi_ 0}{ \sqrt{ 6}M} \right) \Bigg( \dot{ \theta}_ 0 \delta \dot{ \theta} \\
    - \frac{ 1}{ a^ 2} \partial_ i \theta_ 0 \partial^ i \delta \theta \Bigg) + \frac{ 1}{ \lambda} \frac{ d^ 2 V \left( \varphi \right)}{ d \varphi^ 2}|_{ \varphi_ 0} \delta \varphi = 0 \ ,\\
    \\
    \delta \Ddot{ \theta} + 3 H \delta \Dot{ \theta} + \frac{ \Dot{ \lambda}}{ \lambda} \delta \Dot{ \theta} - \frac{ 1}{ a^ 2} \partial_ i^ 2 \delta \theta + \sqrt{ \frac{ 2}{ 3M^2}} \coth \left( \frac{ \varphi_ 0}{ \sqrt{ 6}M} \right) \\
    \times \left( \Dot{ \varphi}_ 0 \delta \Dot{ \theta} - \frac{ 1}{ a^ 2} \partial_ i \varphi_ 0 \partial^ i \delta \theta + \delta \dot{ \varphi} \dot{ \theta}_ 0
    - \partial_ i \delta \varphi \partial^ i \theta_ 0 \right)\\
    - \frac{ 1}{ 3M} \text{csch}^ 2 \left( \frac{ \varphi_ 0}{ \sqrt{ 6}M} \right) \delta \varphi \left( \dot{ \varphi}_ 0 \dot{ \theta}_ 0 - \partial_ i \varphi_ 0 \partial^ i \theta_ 0 \right) = 0 \ ,
    \end{cases}
    \label{perturbed eom}
\end{align}
where we ignore the perturbation of $\lambda$. From the constraint equation \eqref{mimetic constraint multiple field}, the perturbed mimetic constraint is
\begin{align}
    \Dot{ \varphi}_ 0 &\delta \Dot{ \varphi} +6 \sinh^ 2 \left( \frac{ \varphi_0}{ \sqrt{ 6}M} \right) \Dot{ \theta}_ 0 \delta \Dot{ \theta} \nonumber\\
    &+ \frac{\sqrt{ 6}}{M} \sinh \left( \frac{ \varphi_0}{ \sqrt{ 6}M} \right) \cosh \left( \frac{ \varphi_0}{ \sqrt{ 6}M} \right) \delta \varphi \Dot{ \theta}_ 0^ 2 = 0\ .
    \label{perturbed mimetic constraint}
\end{align}
where we considered the homogeneity and isotropy of the background fields. 

\subsection{the inflation epoch}
\label{inflation perturbation}
At the inflation epoch, from the slow-roll condition \eqref{slow roll} one can obtain the equation of motion for $\delta\theta$: 
\begin{align}
    \delta \Ddot{ \theta} + 3 H \delta \Dot{ \theta} +\frac{k^2}{a^2}\delta\theta - \sqrt{ \frac{ 2}{ 3}} \frac{ \dot\varphi}{ M} \delta \Dot{ \theta}= 0\ .
    \label{eom delta theta inflation}
\end{align}
Note that the last term comes from the constraint equation. In our case where $\dot\phi\simeq -\sqrt{2}M^2$ (see Eq. \eqref{dotvarphi}), so $(\dot\varphi/ M) \delta \Dot{ \theta}\simeq \delta \Dot{ \theta}$. As long as we require $M\ll H$, namely the energy scale of the scalar field is much less than the Hubble parameter, the last term can be ignored. Thus Eq. \eqref{eom delta theta inflation} will be the same as the perturbation equation of usual curvaton field. On the other hand,
due to the fact that the mimetic constraint requires the perturbations of both field should be non-vanishing, we also consider the perturbations of $\varphi$, unlike the usual case where we simple neglect the perturbations of inflaton. The perturbed equation of motion of $\varphi$ has a similar equation as $\delta \theta$ at super-horizon scale where $k^2\ll H^2$ again,
\begin{align}
    \delta \Ddot{ \varphi} + 3 H \delta \Dot{ \varphi} + \frac{ V^{ \prime \prime}( \varphi)}{ \lambda} \delta \varphi= 0\ ,
    \label{equation of motion delta varphi inflation}
\end{align}
where $V^{ \prime \prime} \equiv d^ 2 V( \varphi)/ d \varphi^ 2$. Similarly, under the slow-roll condition $V''/H^2\ll 1$, the last term can be ignored. Therefore, we obtain the solutions of $\delta\varphi$ and $\delta\theta$ as:
\begin{align}
    \delta\varphi\simeq\delta\theta\simeq\frac{H}{2\pi}~.
    \label{perturvation of two fields}
\end{align}

\subsection{matter-dominated epoch}
\label{MD perturbation}
In matter dominated universe, the perturbed equations of motion of $\delta \varphi$ and $\delta \theta$ are both wave equations where $3H$ term is cancelled by $\Dot{ \lambda} / \lambda$ term, as demonstrated in Sec. \ref{MD background}. Moreover, taking into account the assumption \eqref{assumptions md},  we can obtain the perturbed equations of motion in matter dominated universe from eq \eqref{perturbed eom} as
\begin{align}
    \delta \Ddot{ \varphi} + \frac{ k^ 2}{ a^ 2} \delta \varphi + \frac{ 2M^4}{ \varphi_ 0^ 2} \delta \varphi + 2 \frac{ \dot{ \varphi_ 0}}{ \varphi_ 0} \delta \dot{ \varphi} + \frac{ 1}{ \lambda} V^{ \prime \prime} \delta \varphi = 0\ ,
    \label{eom delta varphi md}
\end{align}
\begin{align}
    \delta \Ddot{ \theta} + \frac{ k^ 2}{ a^ 2} \delta \theta + 2 \frac{ \dot{ \varphi}_ 0 \dot{ \theta}_ 0}{ \varphi_ 0^ 2} \delta \varphi + \frac{ 2}{ \varphi_ 0}( \delta \dot{ \varphi} \dot{ \theta_ 0} + \dot{ \varphi}_ 0 \delta \dot{ \theta}) = 0\ .
    \label{eom delta theta md}
\end{align}
Furthermore, taking the mimetic constraint \eqref{mimetic constraint multiple field}, perturbed constraint \eqref{perturbed mimetic constraint}, and the solutions \eqref{solution varphi md} and \eqref{solution theta md} into the equations of motion of $\delta \varphi$ \eqref{eom delta varphi md}, it becomes
\begin{align}
    \frac{ d^ 2}{ d t^ 2}( \varphi_ 0 \delta \varphi) + \left( \frac{ k^ 2}{ a^ 2} + \frac{ 3}{ 2} m_{ eff}^ 2 \right) \varphi_ 0 \delta \varphi = 0\ .
    \label{solvable perturbed equation delta varphi md}
\end{align}
Since $m_{ eff} \ll H\lesssim k^2/a^2$, the effective mass term in the above equation can be omitted. Therefore solution of $\delta \varphi$ is
\begin{align}
    \delta \varphi = \frac{ 1}{ \varphi_ 0} \left( D_ + e^{ i k \tau} + D_ - e^{ - i k \tau} \right)\ ,
    \label{solution delta varphi md}
\end{align}
where $\varphi_ 0 \propto t^{ 1 / 2}$. Since the perturbations in this stage are inherited from those in the inflation stage, the coefficients $D_{\pm}$ is determined by the previous solutions. From Eq. \eqref{eom delta theta md} the solution of $\delta \theta$ with $C_\theta = A m_{ eff} / 2 \sqrt{ 3}$ in \eqref{solution theta md} is (see the Appendix \ref{app} for detailed calculations)
\begin{align}
    \delta \theta = \frac{ \sqrt{ 3}}{ 3 A m_{ eff} t} \left( D_{+} e^{ i k \tau} + D_{-} e^{ - i k \tau} \right) \ .
    \label{solution delta theta md}
\end{align}

The perturbations can be transferred to curvature perturbation, when either the curvaton field dominates the universe or the curvaton decays into the background, whatever is earlier \cite{Lyth:2001nq,Lyth:2002my}. The curvature perturbation is related to the density perturbations of each component as
\begin{align}
    \zeta=-H\frac{\delta\rho}{\dot\rho}\simeq \frac{1}{3}\left(\frac{\rho_\varphi}{\rho_{tot}}\delta_\varphi+\frac{\rho_\theta}{\rho_{tot}}\delta_\theta\right)~,
\end{align}
where we define the density contrast of $\varphi$ and $\theta$ as:
\begin{align}
    \delta_\varphi\equiv\frac{\delta\rho_\varphi}{\rho_\varphi}~,~\delta_\theta\equiv\frac{\delta\rho_\theta}{\rho_\theta}~.
\end{align}
Note that for $\varphi$ field, the energy density is mainly contributed by its potential energy, therefore we have $\rho_ \varphi \propto m_{eff}^2 \varphi_ 0^ 2$ and $\delta \rho_\varphi \propto m_{eff}^2\langle \varphi_ 0 \delta \varphi\rangle$, which gives $\delta_\varphi \propto 1 / t$ (Gaussian part) or $1 / t^2$(non-Gaussian part). While for $\theta$ field with no potential, the energy density is mainly contributed by its kinetic energy. Therefore one has $\rho_ \theta \propto \varphi_0^2\dot{ \theta}_ 0^ 2 \sim C_ \theta^ 2 / \varphi_0^2$ and $\delta \rho_\theta \propto \varphi_ 0^ 2\langle \dot{ \theta}_ 0 \delta \dot{ \theta} \rangle =\varphi_ 0^ 2 \sqrt{\dot{ \theta}_ 0^ 2 \langle \delta \theta^ 2\rangle}$, making $\delta_\theta$ time-independent \footnote{Here we used the approximation $\langle\delta\dot\theta^2\rangle\sim\langle\delta\theta^2\rangle$ for oscillating solution of $\delta\theta$.}. Since the $\delta_\varphi$ will dilute as time goes while $\delta_\theta$ does not, and $\varphi_0^2\simeq m_{eff}t\ll 1$ giving $\rho_ \varphi \ll \rho_ \theta $, the term containing $\delta_\theta$ will dominate over the other.

To be more specific, with $\mathcal{ P} = k^ 3| \delta \theta_k|^ 2 / 2 \pi^ 2$, we have the re-entering power spectrum with its initial value correspondence with the value at the end of inflation

\begin{align}
    P_{\theta,md} = P_ {\theta i} \frac{ t_ { decay}^ 2}{ t^ 2}\ ,
    \label{power spectrum delta theta md}
\end{align}
where $t_{ decay}$ is the physical time of the almost total decay of $\varphi$. Following \cite{Lyth:2001nq}, before working out the density contrast $\delta$, we should calculate $\langle\delta \theta^ 2\rangle$ firstly and compare it with $\dot{ \theta}^ 2$ to figure out the feature of power spectrum:
\begin{align}
    \langle\delta &\theta^ 2\rangle = \int_{ k_{ min}}^{ k_{ max}} P_ {\theta, md}( k, t) \frac{ dk}{ k} \nonumber\\
    &\simeq \frac{ H_i^ 2}{ 16 \pi^ 2 M_ P^ 2 \varepsilon_ k} \frac{ t_{ decay}^ 2}{ t^ 2} \text{ln} \left( \frac{ k_{ max}}{ k_{ min}} \right) \ .
    \label{expected delta theta square md}
\end{align}
Thus the Gaussian and non-Gaussian power spectrum are both time independent. The Gaussian power spectrum of curvature perturbation is 

\begin{align}
    P_ \zeta = r^ 2 \frac{H_i^ 2}{ 2 \pi^ 2 M_ P^ 2 \epsilon \dot{ \theta}_ i^ 2}
    \label{power spectrum curvation perutrbation}
\end{align}
where $r\equiv\rho_\theta/\rho_{tot}$ is the energy fraction of curvaton-like field $\theta$ and $\dot{ \theta}_ i$ is the initial value of $\dot{ \theta}$ after preheating as a constant.

The converted curvature perturbation from our curvaton-like field $\theta$ is required to conform to \textit{Planck 2018}. Considering the assumption that the value of effective mass of $\varphi$ is quite small, as $m_{ eff} \ll H$, we point out that the inflationary energy scale is large enough to enable the hot big-bang universe. On the other hand, although the contribution in CMB from $\delta \varphi$ is negligible, what's interesting is that {due to the mimetic constraint as well as the coupling,} we do have fluctuation on inflaton-like field $\varphi$ at inflationary era, which is expected to be tested with the future accuracy of observations.

\section{Conclusion}

In this paper, we study a kind of modified gravity inflation model with inflaton driven by the mimetic gravity field. In such kind of models, due to the constraint condition, the evolution of energy density is solely determined by the potential of inflaton field $\varphi$. Due to the design of the potential, our model can connect the early-time inflation with the late-time matter domination. However, since the single-field mimetic inflation model has problem with quantum fluctuations, we refer to multi-field mimetic inflation model, and in order to explain how adiabatic (curvature) perturbation is generated in multi-field case, we apply the curvaton mechanism, where one of the fields is interpreted as a curvaton field.


To be more precise, we choose the metric of field space as the T-model of $\alpha$-attractor inflation, which keeps the conformal invariance. We have calculated the background and perturbation solutions for both inflation and the following matter-dominant era, where we assume that the reheating process occurs instantaneously. In the inflation era, the inflaton field evolves with a nearly constant velocity $\dot\varphi\simeq-\sqrt{2}M^2$, while the curvaton field nearly remains static due to the mimetic constraint. In the matter-dominant era, the inflaton field oscillates around the minimum of its potential with small effective mass, while the massless curvaton field evolves as $\dot\theta\sim\varphi^{-2}$. To satisfy the constraint equation as well, we also include another field that works as the decay product of both fields. For the perturbation part, since there is a coupling as well as the constraint equation that links the two fields ($\varphi$ and $\theta$), we have to consider perturbations from both inflaton and curvaton. By analytical calculations, we show that these two fields behave similarly as shown in eq. \eqref{perturvation of two fields}. However, one cannot use the slow-roll condition after inflation, thus the differences between inflaton field and curvaton field will evolve differently in MD, for which our calculation shows that the curvature perturbation induced by the isocurvature perturbation is obtained.

Here, we emphasize the importance of our theoretical construction combined with curvaton mechanism and mimetic gravity. Our work suggests that although in mimetic gravity the constraint equation hides the kinetic term of the adiabatic perturbation mode in the action, so as to make it seemingly non-propagate, it actually can be produced from the perturbation of curvaton field, which is mainly isocurvature in the curvaton mechansim, so that one might not need to worry about this non-propagating issue. Moreover, if one introduces the coupling between the two fields, the perturbation of the inflaton field will also appear, which can contribute to the adiabatic perturbations as well. Although in this paper we take a specific example for the illustration, one is free to have it extended to more general cases, such as in the cuscuton cosmological framework \cite{Afshordi:2006ad, Afshordi:2007yx, HosseiniMansoori:2022xnq}.

There still remains lots of questions that needs to be done. Firstly, we have ignored the reheating process, which may have effects on the evolution of perturbations after inflation. Furthermore, from the observational perspective, we could constrain the gravitational waves as well as parameters $r$ which tests its validity for the non-Gaussianities. Finally, we could apply our method to investigate the primordial black hole generation, effects on Big Bang Nucleosynthesis, etc. We postpone these investigations to an upcoming work.


\begin{acknowledgments}
We are grateful to the cosmology group in Central China Normal University (especially Taishi Katsuragawa and Ze Luan) for useful discussions. T.Q. is supported by the National Key Research and Development Program of China under grant NO. 2021YFC2203100, and the National Science Foundation of China under grants no. 11875141. L.L. is funded by the National Science Foundation of China under grant NO. 12165009. 
\end{acknowledgments}

\appendix
\section{Curvature perturbation in matter dominated universe}
\label{app}
Here we consider a little bit more on the case where the decaying product $\psi$ is also included in the constraint equation in matter-dominated epoch, as was discussed in Sec. \ref{MD background}. From Eq. \eqref{explicit eom} The equation of motion for $\varphi$ is
\begin{align}
    \Ddot{ \varphi}_ 0 - \varphi_ 0 \dot{ \theta}_ 0^ 2/M^2 + V_{, \varphi}/\lambda = 0 \ ,
    \label{app eom md}
\end{align}
where the mimetic constraint equation then gives
\begin{align}
    \dot{ \theta}^ 2= \frac{M^2}{ \varphi_ 0^ 2}( 2M^4- \dot{ \psi}^ 2 - \dot{ \varphi}_ 0^ 2) \ .
    \label{app mimetic constraint}
\end{align}
As has been demonstrated, thanks to the $\psi$ field, there will be one more degree of freedom in the constraint equation. Therefore it can be made consistent with the solution of equation of motions of $\varphi$ and $\theta$.
If we construct the potential $V(\varphi)$ as
\begin{align}
    V / \lambda = m_{ eff}^ 2 \varphi_ 0^ 2 / 4 +( 2M^4 - \dot{ \psi}^ 2) \text{ln}( \varphi_ 0 / \varphi_ 1) \ ,
    \label{app potential md}
\end{align}
then the $2M^4 - \dot{ \psi}^ 2$ part in equation of motion of $\varphi$ in Eq. \eqref{app eom md} is cancelled. We still can obtain the equation
\begin{align}
    \frac{ d^ 2}{ dt^ 2}( \varphi_ 0^ 2) + m_{ eff}^ 2 \varphi_ 0^ 2 = 0 \ ,
    \label{app equation varphi square}
\end{align}
and get the solution $\varphi_ 0 = \sqrt{ A} \sin^{ 1 / 2}( m_{ eff} t) \simeq \sqrt{ A m_{ eff} t}$. Therefore the background solution of $\varphi$ and $\theta$ is not affected. On the other hand, assuming $\delta \lambda = 0$ and $\delta \psi = 0$ while applying Eq. \eqref{assumptions md}, we can obtain the perturbed equations of motion
\begin{align}
\begin{cases}
    \delta \Ddot{ \varphi} + \frac{ k^ 2}{ a^ 2} \delta \varphi - \left( \frac{ \varphi_ 0^ 2}{ 6M^2} + 1 \right)\frac{ \dot{ \theta}_ 0^ 2}{ M^ 2} \delta \varphi - \frac{2\varphi_ 0 \dot{ \theta}_ 0 }{M^2} \delta \dot{ \theta}\\
    + \frac{ 1}{ \lambda} V_{, \varphi \varphi} \delta \varphi = 0\\
    \\
    \delta \Ddot{ \theta} + \frac{ k^ 2}{ a^ 2} \delta \theta + \frac{ 1}{ 3} \left(M^2 - \frac{ 6}{ \varphi_ 0^ 2} \right) \dot{ \varphi}_ 0 \dot{ \theta}_ 0 \delta \varphi + 2 \frac{ \dot{ \theta}}{ \varphi} \delta \dot{ \varphi} + 2 \frac{ \dot{ \varphi}}{ \varphi} \delta \dot{ \theta} = 0
\end{cases}
\label{app perturbed eom md}
\end{align}
Since $\varphi_ 0^ 2/M^2 \ll 1$, using Eqs. \eqref{app mimetic constraint} and \eqref{app potential md}, we can derive the equation of $\varphi_ 0 \delta \varphi$ as
\begin{align}
    \frac{ d^ 2}{ dt^ 2}( \varphi_ 0 \delta \varphi) + \left( \frac{ k^ 2}{ a^ 2} + \frac{ 1}{ 2} m_{ eff}^ 2 - \frac{ \dot{ \varphi}_ 0^ 2}{ \varphi_ 0^ 2} - \frac{ \Ddot{ \varphi}_ 0}{ \varphi_ 0} \right) \varphi_ 0 \delta \varphi = 0 \ ,
\end{align}
where $\dot{ \varphi}_ 0^ 2 / \varphi_ 0^ 2 = - \Ddot{ \varphi}_ 0 / \varphi_ 0 = 1 / 4 t^ 2$. Since $m_{ eff}^ 2 \ll H^ 2$, the $m_{ eff}^ 2$ term can be omitted, thus the solution of $\delta \varphi$ is again Eq. \eqref{solution delta varphi md}. Moreover, considering the perturbed mimetic constraint
\begin{align}
    - \dot{ \theta} \delta \dot{ \theta} \simeq \frac{ M^ 2}{ \varphi_ 0^ 2}( \dot{ \varphi}_ 0 \delta \dot{ \varphi}+ \frac{ \varphi_ 0 \delta \varphi}{ M^ 2} \dot{ \theta}^ 2) \ 
\end{align}
and the solution of $\dot{ \theta}_ 0 = C_\theta / \varphi_ 0^ 2$, the last three term in equation of $\delta \theta$ in Eq. \eqref{app perturbed eom md} becomes
\begin{align}
    - 2 \frac{ C_\theta \dot{ \varphi}_ 0}{ \varphi_ 0^ 4}\delta \varphi + 2 \frac{ C_\theta}{ \varphi_ 0^ 3} \delta \dot{ \varphi} - \frac{ 2}{ C_\theta} \frac{ \dot{ \varphi}_ 0}{ \varphi_ 0}( \dot{ \varphi}_ 0 \delta \dot{ \varphi} + \varphi_ 0 \dot{ \theta}^ 2 \delta \varphi) \ ,
\end{align}
where
\begin{align}
    - \frac{ 2 C_\theta \dot{ \varphi}_ 0}{ \varphi_ 0^ 4} &\simeq \frac{ C_\theta}{( A m_{ eff})^{ 3 / 2} t^{ 5 /2}} \ ,\\
    2 \frac{ C_\theta}{ \varphi_ 0^ 3} \delta \dot{ \varphi} &\simeq \frac{ C_\theta}{( A m_{ eff} t)^{ 3 / 2}}\left( \pm i \frac{ 2 k}{ a} - \frac{ 1}{ t} \right) \delta \varphi \ ,\\
    - \frac{ 2}{ C_\theta} \frac{ \dot{ \varphi}_ 0^ 2}{ \varphi_ 0} \delta \dot{ \varphi} &\simeq - \frac{ \sqrt{ A m_{ eff}}}{ 4 C_\theta t ^{ 3 / 2}} \left( \pm i \frac{ 2 k}{ a} - \frac{ 1}{ t} \right) \delta \varphi \ ,
\end{align}

Without the third line, $\delta \theta$ will evolves time-independently after it gets close to the horizon, which is not an acceptable solution for curvature perturbation. This point requires $12 C_\theta ^ 2 = A^ 2 m_{ eff}^ 2$, so that the perturbed equation \eqref{app perturbed eom md} becomes
\begin{align}
    \delta \Ddot{ \theta} + \frac{ k^ 2}{ a^ 2} \delta \theta - \frac{ 2 \sqrt{ 3}}{ 3 A m_{ eff}} \left[\frac{ 1}{ t^ 3} \mp i \frac{ k}{ a} \frac{ 1}{ t^ 2} \right] D_{ \pm} e^{ \pm i k \tau} = 0 \ ,
\end{align}
with solution
\begin{align}
    \delta \theta = \frac{ \sqrt{ 3}}{ 3 A m_{ eff} t} D_{ \pm} e^{ \pm i k \tau} \approx \frac{ \sqrt{ 3}}{ 3 \varphi_ 0^ 2} D_{ \pm} e^{ \pm i k \tau} \ .
    \label{app solution delta theta md}
\end{align}
which has an oscillating behavior with the amplitude proportional to time inverse.

\end{document}